\documentclass[a4paper,12pt]{article}
\def\beq{\begin{equation}}
\def\eeq{\end{equation}}
\def\bea{\begin{eqnarray}}
\def\eea{\end{eqnarray}}
\def\beax{\begin{eqnarray*}}
\def\eeax{\end{eqnarray*}}
\def\vf{\varphi}
\def\half{\frac{1}{2}}
\def\quarter{\frac{1}{4}}
\begin{document}
\title{On the Path-Integral Derivation of the Anomaly for the Hermitian
Equivalent of the Complex $PT$-Symmetric Quartic Hamiltonian}
\author{H.~F.~Jones${}^1$, J.~Mateo${}^2$ and R.~J.~Rivers${}^1$\\
${}^1${\it Physics Department, Imperial College, London SW7, UK}\\
${}^2${\it Departamento de F\'isica Te\'orica, At\'omica y \'Optica, }\\
{\it Facultad de Ciencias,  E-47011, Valladolid, Spain}}
\date{11/10/06}
\maketitle \centerline{\bf Abstract}
{It can be shown using operator techniques that the non-Hermitian $PT$-symmetric
quantum mechanical Hamiltonian with a ``wrong-sign" quartic potential $-gx^4$ is
equivalent to a Hermitian Hamiltonian with a positive quartic potential together with
a linear term. A na\"{\i}ve derivation of the same result in the path-integral approach
misses this linear term. In a recent paper by Bender et al. it was pointed out that this
term was in the nature of a parity anomaly and a more careful, discretized treatment of
the path integral appeared to reproduce it successfully. However, on re-examination of this
derivation we find that a yet more careful treatment is necessary, keeping terms that were
ignored in that paper. An alternative, much simpler derivation is given using the additional
potential that has been shown to appear whenever a change of variables to curvilinear coordinates
is made in a functional integral. }\\


\thispagestyle{empty}

\section{Introduction}
Hamiltonians of the form
\bea\label{HN}
H=\half p^2-g (ix)^N,
\eea
which are are $PT$-symmetric, but not Hermitian,
have been shown\cite{BB, Dorey} to have a real, positive
spectrum for $N\ge2$. For $N<4$ the energy eigenvalue problem can
be posed on the real $x$-axis, but for $N\ge4$ it must instead be imposed
in the complex $x$-plane, along a contour which asymptotically remains
within the Stokes wedges of the Hamiltonian. Such is indeed the case
for $N=4$, when the potential on the real axis is an ``upside down"
quartic. To that end we rewrite Eq.~(\ref{HN}) for $N=4$ in terms of
the complex variable $z$:
\bea\label{H4}
H=\half p_z^2 -gz^4.
\eea

A general framework encompassing $PT$-symmetric Hamiltonians was given by
Mostafazadeh\cite{AM}, who showed that such Hamiltonians were related
by a similarity transformation to an equivalent Hermitian Hamiltonian
$h$ possessing the same spectrum. In only a few cases is it possible to
construct $h$ explicitly, but this was done recently for Eq.~(\ref{H4}) in Ref.~\cite{JM},
rediscovering an earlier result of Buslaev and Grecchi\cite{BG}. The calculation
was performed using the particular real parametrization
\bea\label{param}
z=-2i\sqrt{1+ix}
\eea
of the contour in the
complex plane on which Eq.~(\ref{H4}) was defined, and using operator techniques
to perform the similarity transformation to $h$, whose form is ($\alpha\equiv16g$)
\bea\label{h}
h=\half p_y^2+\quarter \alpha y^4 -\sqrt{\frac{\alpha}{8}}y.
\eea
We should emphasize that this latter form is obtained only after a Fourier transform,
which means that $y$ is really a momentum rather than a position, as will become
apparent in the path-integral treatment. When a proper dimensional analysis is made\cite{An1},
the linear term turns out to be proportional to $\hbar$, which means that it is
in the nature of an anomaly.

In fact we will present various path-integral treatments, of various levels of
sophistication. In the next section we give the simplest of these, a continuum  version, which fails
to reproduce the anomaly. Realizing that for a general change of variables in a functional
integral there is an additional term, $\Delta V$, to be included over and above the functional determinant,
we show, in Section~\ref{sec:Lee}, how $\Delta V$ is traded for the anomaly in an intermediate functional integration.
In Section 4 we give the discretized version of this calculation, in order to provide a benchmark
for comparison with the results obtained in Section 5 by two variants of the discretized calculation
of Ref.~\cite{An1}. In effect this latter section gives an alternative derivation of $\Delta V$, as a consequence
of the mismatch between the functional determinant and the coefficient of the kinetic term when our
particular change of variable is made.

\section{Na\"{\i}ve Continuum Treatment}

The original path integral is the single Euclidean functional integral
\bea
Z=\int_C [D\psi] \exp \left\{-\int dt\left[\half\dot{\psi}^2-g\psi^4\right]\right\},
\eea
where the subscript $C$ is to remind us that the integrals are to be performed on an
appropriate curve in the complex $\psi$ plane. Making the change of variable
\bea\label{cv}
\psi=-2i\surd(1+i\,\vf),
\eea
in analogy with Eq.~(\ref{param}),
we obtain
\bea\label{Znaive}
Z=\int \frac{[D\vf]}{{\rm Det}\surd(1+i\,\vf)} \exp \left\{-\int dt\left[\half\frac{\dot{\vf}^2}{1+i\,\vf}
-\alpha(1+i\,\vf)^2\right]\right\},
\eea
where the $\vf$ integrals are now along the real axis.
We now rewrite this as a double functional
integral in $\vf$ and $\pi$ by means of the identity
\bea\label{Pi1}
\frac{1}{{\rm Det}\surd(1+i\,\vf)}=\frac{1}{N}\int [D\pi] \exp \left\{ -\int dt\,\half
(1+i\,\vf)\left(\pi-i\frac{\dot{\vf}}{1+i\,\vf}\right)^2\right\},
\eea
where $N$ is an appropriate normalization constant.
This leads to the expression
\bea
Z=\frac{1}{N}\int [D\vf][D\pi] \exp \left\{-\int dt\left[\half(1+i\,\vf)\pi^2-i\pi\dot{\vf}-\alpha(1+i\,\vf)^2
\right]\right\}.
\eea
Replacing $\pi\dot{\vf}$ by -$\dot{\pi}\vf$ under the $t$ integration, we have an exponent that is
quadratic in $\vf$, so the $\vf$ functional integration can be performed.
The result, after applying the compensating rescalings
$\vf\to \vf/\sqrt{2\alpha}$ and $\pi\to \pi\sqrt{2\alpha}$, is
\bea
Z=\int[D\pi] \exp \left\{-\int dt\left[\half\dot{\pi}^2-\sqrt{2\alpha}\dot{\pi}\left(1-\frac{1}{2}\pi^2\right)+
\frac{\alpha}{4}\pi^4
\right]\right\}.
\eea
The middle term in the integrand is a perfect derivative, and so may be discarded, leaving
the equivalent Lagrangian $\ell$
in the exponent, written in terms of $\pi$:
\bea\label{eq:naive}
Z=\int[D\pi] \exp \left\{-\int dt\left[\half\dot{\pi}^2+
\frac{\alpha}{4}\pi^4
\right]\right\},
\eea
\emph{except} that the linear term is missing. The simple classical calculation we have just performed is unable
to obtain this term. Note that the Fourier transform of the Schr\"{o}dinger treatment occurs here naturally,
since $Z$ is expressed in terms of the momentum variable $\pi$.

\section{Correct Continuum Treatment}\label{sec:Lee}
It is shown in various standard books on functional integration, for example \cite{Lee,CD}, that when a general change
of variables, such as that of Eq.~(\ref{cv}), is made, an additional potential term $\Delta V$ must
be included in the Lagrangian. Ultimately this term, which is actually of order $\hbar^2$, is derived from the
discretized form of the functional integral when the particular form of discretization
\bea
\dot{\vf_n}&\equiv& \frac{1}{a}(\vf_{n+1}-\vf_n)\\
\bar{\vf_n}&\equiv&\half (\vf_{n+1}+\vf_n)
\eea
is adopted, where $a$ is the lattice spacing, corresponding to Weyl ordering in the operator treatment.
The importance of this prescription was indeed emphasized in Ref.~\cite{An1}.

The general form of $\Delta V$ for a change of variable from $\psi$ to $\vf$ is\cite{Lee}
\bea\label{Vcgen}
\Delta V=\frac{1}{8}\left[\frac{d}{d\vf}\left(\frac{d\vf}{d\psi}\right)\right]^2,
\eea
which for the particular transformation of  Eq.~(\ref{cv}) turns out to be
\bea\label{Vc}
\Delta V=-\frac{1}{32}\frac{1}{1+i\vf}.
\eea
The correct version of Eq.~(\ref{Znaive}) is thus
\bea\label{ZLee}
Z&=&\int \frac{[D\vf]}{{\rm Det}\surd(1+i\,\vf)} \exp \left\{-\int dt\left[\half\frac{\dot{\vf}^2}{1+i\,\vf}
-\frac{1}{32}\frac{1}{1+i\vf}-\alpha(1+i\,\vf)^2\right]\right\}.\nonumber\\
&&
\eea
It is now possible to write down a variant of the Gaussian identity (\ref{Pi1}),
\bea\label{Pi2}
\frac{1}{{\rm Det}\surd(1+i\,\vf)}=\frac{1}{N}\int [D\pi] \exp \left\{ -\int dt\,\half
(1+i\,\vf)\left(\pi-\frac{i\dot{\vf}+\quarter}{1+i\,\vf}\right)^2\right\},
\eea
which serves to cancel the $\Delta V$ term as well as the kinetic term. In turn it introduces two additional
terms in the exponent: (i) a term in $i\dot{\vf}/(1+i\vf)$, which is a perfect derivative and so can be
discarded under the $t$ integration, and (ii) the anomaly $\int dt\ \pi/4$.

Finally, after rescaling as before, the corrected version of Eq.~(\ref{eq:naive}) is
\bea
Z=\int[D\pi] \exp \left\{-\int dt\left[\half\dot{\pi}^2-\sqrt{\frac{\alpha}{8}}\pi+
\frac{\alpha}{4}\pi^4
\right]\right\},
\eea

\section{Discretized Version}
In this section we will go through the discretized version of the previous calculation, in
order to provide a standard discretized formula with which we can compare the results of the (corrected)
calculation of Ref.~\cite{An1} and another calculation whereby the kinetic term is expanded in powers of
the lattice spacing $a$.

In place of Eq.~(\ref{ZLee}) we have
\bea
Z=\prod_n\int \frac{d\vf_n}{\sqrt{1+i\bar{\vf}_n}} \exp\left\{-a\left[\half\frac{\dot{\vf}_n^2}{1+i\bar{\vf}_n}
-\frac{1}{32}\frac{1}{{1+i\bar{\vf}_n}}-\alpha(1+i\bar{\vf}_n)^2   \right]\right\}.
\eea
The Gaussian identity we will use is
\bea
\frac{1}{\sqrt{1+i\bar{\vf}_n}}=\frac{1}{N}\int d\bar{\pi}_n e^{-\half\lambda(\bar{\pi}_n-B)^2},
\eea
where $\lambda=a(1+i\bar{\vf}_n)$ and $B=(i\dot{\vf}_n+\frac{1}{4})/(1+i\bar{\vf}_n)$.

Written out in full this is
\beax
\frac{1}{\sqrt{1+i\bar{\vf}_n}}&=&\frac{1}{N}\int d\bar{\pi}_n\exp\left\{-a\left[\half \bar{\pi}_n^2(1+i\bar{\vf}_n)
-i\bar{\pi}_n\dot{\vf}_n- \frac{1}{4}\bar{\pi}_n\right.\right.\\
&&\left.\left.\hspace{2.7cm}+\quarter\frac{i\dot{\vf}_n}{1+i\bar{\vf}_n}-\half\frac{\dot{\vf}_n^2}{1+i\bar{\vf}_n}+\frac{1}{32}\frac{1}{1+i\bar{\vf}_n}
\right]\right\}.
\eeax
Neglecting the term in $i\dot{\vf}_n/(1+i\bar{\vf}_n)$ because the identity
\beq
\log\left(\frac{1+i\vf_{n+1}}{1+i\vf_n}\right)=\frac{ia\dot{\vf}_n}{1+i\bar{\vf}_n}
+O(a^3)
\eeq
shows it to be a perfect difference up to a correction of order $a^3$, we obtain
 \beq\label{bench}
 Z=\frac{1}{N}\prod_n\int d\bar{\pi}_n d\vf_n \exp\left\{-a\left[ \half(1+i\bar{\vf}_n)\bar{\pi}_n^2-i\bar{\pi}_n\dot{\vf}_n-
 \frac{1}{4}\bar{\pi}_n-\alpha(1+i\bar{\vf}_n)^2)  \right]\right\}
\eeq
Thus the $\Delta V$ term  has been cancelled,  and we are left with the anomaly $\frac{1}{4}a\bar{\pi}_n$
in the exponent.
Now $\bar{\pi}_n\dot{\vf}_n+\bar{\vf}_n\dot{\pi}_n$ is a perfect difference:
\beq
a(\bar{\pi}_n\dot{\vf}_n+\bar{\vf}_n\dot{\pi}_n)=
\pi_{n+1}\vf_{n+1}-\pi_n\vf_n.
\eeq
So now we can ``integrate by parts", in the form $\bar{\pi}_n\dot{\vf}_n\to-\bar{\vf}_n\dot{\pi}_n$.\\
Changing the integration measure
from $\int d\bar{\pi}_n d\vf_n$ to
$\int  d\pi_n d\bar{\vf}_n$, which does not introduce any additional factors, Eq.~(\ref{bench}) becomes
\beax
 Z&=&\frac{1}{N}\prod_n\int d\pi_n d\bar{\vf}_n \exp\left\{-a\left[
 \alpha\left((\bar{\vf}_n-i)+\frac{i}{4\alpha}(\bar{\pi}_n^2+2\dot{\pi}_n)\right)^2\right.\right.\\
 &&\hspace{4cm}\left.\left.+\frac{1}{16\alpha}\left( \bar{\pi}_n^2+2\dot{\pi}_n\right)^2 -\frac{1}{4}\bar{\pi}_n\right]\right\},
 \eeax
 having dropped a perfect difference proportional to $\dot{\pi}_n$.

 Now we rescale: $\bar{\vf}_n\to \bar{\vf}_n/\sqrt{2\alpha}$ and $\pi_n\to \pi_n\sqrt{2\alpha} $ and perform
 the $\bar{\vf}_n$ integration, with the result
 \beax
 Z=\frac{1}{N}\prod_n\int d\pi_n \exp\left\{-a\left[
 \frac{\alpha}{4}\bar{\pi}_n^4+\sqrt{\frac{\alpha}{2}}\bar{\pi}_n^2\dot{\pi}_n+\frac{1}{2}\dot{\pi}_n^2
  -\sqrt{\frac{\alpha}{8}}\bar{\pi}_n \right]\right\}
 \eeax
This is the desired result, provided that we can neglect the term $\bar{\pi}_n^2\dot{\pi}_n$.
The identity
\bea
3a\bar{\pi}_n^2\dot{\pi}_n=\pi_{n+1}^3-\pi_n^3-(\pi_{n+1}-\pi_n)^3/4
\eea
shows that is a perfect difference up to a correction of order $a^3$, so that it can indeed be neglected.
The resulting expression for Z is
\bea
 Z=\prod_n\int d\pi_n \exp\left\{-a\left[
 \frac{1}{2}\dot{\pi}_n^2
  -\sqrt{\frac{\alpha}{8}}\bar{\pi}_n +\frac{\alpha}{4}\bar{\pi}_n^4\right]\right\},
 \eea
the discrete version of Eq.~(\ref{ZLee}).
Equation (\ref{bench}), from which this was derived, will be a reference point for the calculations of the following two sections, which
arise out of the treatment of Ref.~\cite{An1}.

The basis of that treatment was an exact, discretized, treatment of the kinetic term.
In the original functional integral written in terms of $\psi$, the time derivative is
defined as
$\dot\psi_n=(\psi_{n+1}-\psi_n)/a.$
In terms of $\psi_n=-2i\surd(1+i\vf_n)$, this becomes
$\dot\psi_n=\dot{\vf_n}/A_n$,
where
\bea
A_n=\half(\sqrt{1+i\vf_{n+1}}+\sqrt{1+i\vf_n})
\eea
The relation between $A_n$ and $\sqrt{1+i\vf_n}$ can be written exactly as
\bea\label{determinant}
\frac{1}{\sqrt{1+i\vf_n}}=\frac{1}{A_n}\left(1+\frac{ia\dot{\vf_n}}{4A_n\sqrt{1+i\vf_n}}\right)
\eea
In this formulation the anomaly arises because the denominator $A_n$ in the expression for $\dot{\vf}_n$ is not
quite the same as that in the determinant of Eq.~(\ref{determinant}). That is, we have the expression
$\exp[-a\dot{\vf_n}^2/(2A_n^2)]/\surd(1+i\vf_n)$,
and we need to expand one of these in
terms of the other.
\section{Expanding the Kinetic Term}\label{KE}

In this case we take $\lambda=a(1+i\vf_n)$, $x=\bar{\pi}_n$ and $B=i\dot{\vf_n}/(1+i\vf_n)$ in the
Gaussian identity
\bea\label{GI}
\frac{1}{\sqrt{1+i\vf_n}}&=&\frac{1}{N}\int dx \exp\left\{-\half\lambda(x-B)^2\right\}\\
&=&\frac{1}{N}\int d\bar{\pi}_n \exp\left\{-\half a(1+i\vf_n)\bar{\pi}_n^2
+ia\dot{\vf}_n\bar{\pi}_n+\half\frac{a\dot{\vf}_n^2}{1+i\vf_n}\right\}\nonumber
\eea
Now we need to expand the kinetic term $-\half a\dot{\vf_n}^2/A_n^2$ in terms of $\half\lambda B^2$, namely
$\half\lambda B^2=-\half a\dot{\vf_n}^2/(1+i\vf_n)$.
First write Eq.~(\ref{determinant}) in terms of $B$:
\beax
\frac{1}{\sqrt{1+i\vf_n}}=\frac{1}{A_n}\left(1+\frac{aB}{4A_n}\sqrt{1+i\vf_n}\right)
\eeax
In terms of $R\equiv{\sqrt{1+i\vf_n}}/{A_n}$, this reads
\bea
\frac{1}{R}=1+\frac{1}{4}aBR,
\eea
which is a quadratic equation for $R$, with solution
\beax
R=\frac{2}{1+\sqrt{1+aB}}.
\eeax
So the kinetic term is $-\half a\dot{\vf_n}^2/A_n^2=\half\lambda B^2 R^2=
\half \lambda B^2+\half\lambda B^2 (R^2-1).$\\
Hence
\beq\label{keep}
e^{-\half a\dot{\vf_n}^2/A_n^2}
=e^{\half\lambda B^2}\left[1-\quarter a\lambda B^3 +\frac{a^2}{32}(5\lambda B^4
+\lambda^2 B^6)  +\dots \right]
\eeq
Now we need the Gaussian identities, under ${\large\int} dt \exp\{-\half\lambda(x-B)^2\}$:
\bea
B^3&\equiv& x^3-\frac{3x}{\lambda}, \nonumber\\
B^4&\equiv& x^4-\frac{6x^2}{\lambda}+\frac{3}{\lambda^2},\\
B^6&\equiv& x^6-\frac{15x^4}{\lambda}+\frac{45x^2}{\lambda^2}-\frac{15}{\lambda^3},\nonumber
\eea
to get
\bea\label{wrongK}
e^{-\half a\dot{\vf_n}^2/A_n^2}=e^{\half\lambda B^2}\left(1+\frac{3}{4}a\bar{\pi}_n + O(a^2)\right).
\eea
Note that it is crucial to keep terms in Eq.~(\ref{keep}) which are nominally of order $a^2$ (and
possibly higher). The point is that the Gaussian identities bring in terms of order $1/\lambda$, $1/\lambda^2$
etc., which means that such terms may actually be of order $a$.
The net result so far is that  we appear to have produced the anomaly, but with the wrong coefficient, 3/4 versus 1/4. However,
it is important to realize that after these transformations we are
left with $-\half a(1+i\vf_n)\bar{\pi}_n^2$ in the exponent, as opposed to
the $-\half a(1+i\bar{\vf}_n)\bar{\pi}_n^2$ of Eq.~(\ref{bench}).
The difference between them is
\beax
-\half a(1+i\vf_n)\bar{\pi}_n^2&=&-\half a(1+i\bar{\vf}_n)\bar{\pi}_n^2+\quarter i a^2\dot{\vf}_n \bar{\pi}_n^2\\
&=&-\half a(1+i\bar{\vf}_n)\bar{\pi}_n^2+\quarter\lambda aB \bar{\pi}_n^2
\eeax
In order to implement Gaussian identities with the new kinetic term we need to
write the correction in terms of a new $\lambda$ and $B$, namely
$\bar\lambda= a(1+i\bar{\vf}_n)$ and $\bar{B}= i\dot{\vf}_n/(1+i\bar{\vf}_n)$,
with the same property that $\bar{\lambda}\bar{B}=\lambda  B=ia\dot{\vf}_n$.
The exponential of the additional term $\quarter\bar{\lambda} a\bar{B} x^2$ expands to
\beax
\exp(\quarter\bar{\lambda} a\bar{B} x^2)=1+\quarter\bar{\lambda} a\bar{B}x^2+\frac{1}{32}\bar{\lambda}^2\bar{B}^2x^4+\dots
\equiv 1-\half ax+O(a^2),
\eeax
using the further Gaussian equivalences
\bea
\bar{B}x^2&\equiv&x^3-\frac{2x}{\bar{\lambda}},\nonumber\\
\bar{B}^2x^4&\equiv&x^6-\frac{9x^4}{\bar{\lambda}}+\frac{12x^2}{\bar{\lambda}^2}.
\eea
Thus, the additional term precisely corrects
the coefficient of the anomaly from 3/4 in Eq.~(\ref{wrongK}) to 1/4.
A necessary ingredient for this to work is the lack of a term in $1/\bar{\lambda}^3$ in the
Gaussian equivalence for $\bar{B}^2x^4$.

\section{Expanding the Determinant}\label{detsec}

In this case we take $x=\bar{\pi}_n$, $\lambda=aA_n^2$ and $B=i\dot{\vf_n}/A_n^2$ in the
Gaussian identity
\bea\label{Gauss}
\frac{1}{A_n}&=&\frac{1}{N}\int dx \exp \left\{
-\half\lambda(x-B)^2\right\}\\
&=&\frac{1}{N}\int dx \exp \left\{-\half aA_n^2\bar{\pi}_n^2+ia\dot{\vf}_n\bar{\pi}_n+\frac{a\dot{\vf}_n^2}{2A_n^2}\right\}
\eea

 In fact we can solve Eq.~(\ref{determinant}) for $\sqrt{1+i\vf_n}$ in terms of $A_n$:
\bea
\frac{1}{A_n}=\frac{1}{\sqrt{1+i\vf_n}}\left(1-\frac{ia\dot{\vf_n}}{4A_n^2}\right)
=\frac{1}{\sqrt{1+i\vf_n}}\left(1-\frac{1}{4}aB\right)
\eea
Thus, in terms of the new $B$,
$R\equiv\sqrt{1+i\vf_n}/A_n=1-aB/4$.

So the determinant can be written as
\beq
\frac{1}{\sqrt{1+i\vf_n}}=\frac{1}{A_n}\cdot \frac{1}{R}
=\frac{1}{A_n}\left(1+\frac{1}{4}aB+\frac{1}{16}a^2B^2 +\dots\right)
\eeq
Now we use two further identities under the Gaussian integration of Eq.~(\ref{Gauss}):
\bea
B&\equiv& x \nonumber\\
B^2&\equiv& x^2-\frac{1}{\lambda}
\eea
to obtain
\bea\label{first}
\frac{1}{\sqrt{1+i\vf_n}}
\equiv \frac{1}{A_n}\left(1+\frac{1}{4}ax-\frac{a}{16A_n^2} +O(a^2)\right)
\eea
The second term in $\big(\hspace{1cm}\big)$ correctly gives the anomaly when elevated to the
exponent, but the third term, which was missed in Ref.~\cite{An1}, appears to be an unwanted addition.
The final result is
\beq\label{eqdet}
Z=\frac{1}{N} \prod_n\int d\vf_n d\bar{\pi}_n \exp\left\{-a\left[\half A_n^2 \bar{\pi}_n^2-i\bar{\pi}_n\dot{\vf}_n
-\frac{1}{4}\bar{\pi}_n +\frac{1}{16A_n^2}-\alpha(1+i\vf_n)^2\right]\right\}\nonumber
\eeq

This differs from Eq.~(\ref{bench})
 in two ways: we have a term resembling $\Delta V$, and the coefficient of $\bar{\pi}_n^2$
is $\half A_n^2$ rather than $\half(1+i\bar{\vf}_n)$. The difference between $\vf_n$ and $\bar{\vf}_n$ in the interaction
term is not important.

Now
\beax
aA_n^2\bar{\pi}_n^2=a(1+i\vf_n)\bar{\pi}_n^2+\lambda\bar{\pi}_n^2\left(1-R^2\right).
\eeax
So, since $ia\vf_n=ia(\bar{\vf}_n-\half a \dot{\vf}_n)=ia\bar{\vf}_n-\half\lambda aB$,
\beax
-\half aA_n^2\bar{\pi}_n^2=-\half a(1+i\bar{\vf}_n)\bar{\pi}_n^2+\frac{1}{4}\lambda (aB) \bar{\pi}_n^2
-\half \lambda\bar{\pi}_n^2\left(1-R^2\right).
\eeax
The additional terms expand to
\beax
\frac{1}{4} \lambda\bar{\pi}_n^2 (aB)-\half\lambda\bar{\pi}_n^2\left[\half(aB)-\frac{1}{16}(aB)^2\right]
=\frac{1}{32}\lambda\bar{\pi}_n^2(aB)^2
\eeax
and their exponential to $1+(aB)^2\lambda\bar{\pi}_n^2/32+\dots$.

Again, because we have changed the coefficient of the kinetic term to $\half(1+i\bar{\vf}_n)$ we
strictly need to rewrite $\lambda B^2$ in terms of $\bar{\lambda}$ and $\bar{B}$
according to $\lambda B^2 = (\bar{\lambda}\bar{B}^2)\times (\bar{\lambda}/\lambda)$,
where the correction factor is
\beax
\frac{\bar{\lambda}}{\lambda}=\frac{1+i\bar{\vf}_n}{A_n^2}=1+O(a\bar{B})^2
\eeax
The additional term does not in fact contribute to order $a$.
Thus we need the final equivalence
\bea\label{xsqBsq}
\bar{B}^2x^2&\equiv&x^4-\frac{5x^2}{\bar{\lambda}}+\frac{2}{\bar{\lambda}^2}
\eea
for $\bar{B}^2x^2$ under Gaussian integration, which gives
$1+a^2/(16\bar{\lambda})+O(a^3)$.
When exponentiated this term precisely cancels the $\Delta V$-like term $a/(16A_n^2)$ in
Eq.~(\ref{eqdet}) up to $O(a^2)$.

\section{Conclusions}
The na\"ive functional integral formulation of the operator calculation can be regarded as the
classical result, in that it takes no account of operator ordering, or equivalently of any particular
discretization, and so fails to produce the linear, anomalous term, which is of order $\hbar$.
We have shown an elegant method of producing this term in the continuous formalism, provided we take
account of the additional $\Delta V$ term that has been shown (by careful discretization) to occur whenever
a general change of variables is made.

The calculation of Ref.~\cite{An1} indeed attempted a careful discretization, but missed terms that were
nominally of higher order in the lattice spacing $a$, but were in fact of the same order as the terms kept
because of the particular nature of the Gaussian equivalences, which bring in factors of $O(1/\lambda)=O(1/a)$.
Starting from the recognition that the coefficient of the kinetic term does not exactly match the argument of
the functional determinant we proceeded in two ways, expanding either the kinetic term or the functional determinant.
In both cases we used the discretized version of the corrected continuum calculation as a canonical form with
which to compare our results. The expansion of the kinetic term turned out to be the easier, needing only a
change from the point variable $\vf_n$ to the averaged variable $\bar{\vf}_n$ to obtain agreement. The method
of expanding the determinant used in Ref.~\cite{An1} eventually produced the same result after more lengthy
manipulations, which incidentally showed that the simple passage from Eq.~(57) to Eq.~(58) in Ref.~\cite{An1}
was not correct.

One can ask whether the same phenomenon of trading a $\Delta V$ term for an anomaly by means of
some analogue of Eq.~(\ref{Pi2}) is possible for other changes of variable. The answer seems to
be in the negative: the parametrization of Eq.~(\ref{cv}) seems to be very special, as indeed
it is in other respects.

In general, the coefficient of the kinetic term in $\vf$ starting from a standard kinetic term in
$\psi$ is $\half M$, where
\beq
M=\left(\frac{d\psi}{d\vf}\right)^2,
\eeq
to be compared with the expression for $\Delta V$ in Eq.~(\ref{Vcgen}). For the two to be proportional
we require
\beq
\frac{d\psi}{d\vf}\propto\frac{d}{d\vf}\left(\frac{d\vf}{d\psi}\right),
\eeq
or equivalently $d^2\vf/d\psi^2=\mbox{const}$. This leads back to a relation of the general form of
Eq.~(\ref{cv}) up to a shift in $\psi$.

The main motivation for reformulating the quantum mechanical problem in path-integral terms in
Ref.~\cite{An1} was in order to attempt a generalization to higher dimensions, where there are
indications from the non-Hermitian formulation that the theory is asymptotically free and naturally
possesses a non-vanishing vacuum expectation value, making it an attractive alternative to the standard
Higgs model. It is therefore
natural to ask whether additional potentials of the nature of $\Delta V$, and the consequent
production of an anomaly, occur in dimensions greater than 1. The answer appears to be negative,
at least in the context of dimensional regularization.  This is because algebraic field transformations
lead to $O(\hbar^2)$ terms proportional to $(\delta^{(n)}(0))^2$ in $n$ spatial dimensions,
which vanish in such a regularization scheme. This is consistent with a diagrammatic expansion,
for which no additional terms are necessary \cite{HV}. The reason for their presence in quantum
mechanics versus their absence in field theory has been analyzed in some detail by Salomonson\cite{PS}.

Even though it appears that this particular problem does not occur in field theory, there remain
severe difficulties in carrying out the programme. The most promising approach (3) put forward
in Ref.~~\cite{An1} still contains uncancelled Jacobian factors, which can only be represented
in the Lagrangian by the introduction of several auxiliary fields.\\

\noindent{\large\bf Acknowledgements}\\

This work has been partially supported by SpanishMEC (ProjectMTM2005-
09183) and Spanish JCyL (Excellence Project VA013C05). We are grateful
to C.~M. Bender and J.~M. Cerver\'o for helpful discussions.

\end{document}